# Steganography and Steganalysis: Different Approaches

Soumyendu Das
Information Security Consultant
Kolkata, India, soumyendu.das@gmail.com

Subhendu Das
STQC IT Services, Kolkata, India, subhendu.das@gmail.com

Bijoy Bandyopadhyay
Institute of Radio physics & Electronics,
University of Calcutta, Kolkata, India, bbandy@vsnl.com

Sugata Sanyal
Tata Institute of Fundamental Research
Mumbai, India, sanyal@tifr.res.in

**Abstract**

Steganography is the technique of hiding confidential information within any media. Steganography is often confused with cryptography because the two are similar in the way that they both are used to protect confidential information. The difference between the two is in the appearance in the processed output; the output of steganography operation is not apparently visible but in cryptography the output is scrambled so that it can draw attention. Steganlysis is process to detect of presence of steganography. In this article we have tried to elucidate the different approaches towards implementation of steganography using 'multimedia' file (text, static image, audio and video) and Network IP datagram as cover. Also some methods of steganalysis will be discussed.

**Key words**

Steganography, Steganalysis, Discrete Cosine Transformation (DCT), Ipv4 header, IP datagram fragmentation.

## 1. Introduction

The objective of steganography is to hide a secret message within a cover-media in such a way that others cannot discern the presence of the hidden message. Technically in simple words "steganography means hiding one piece of data within another".

Modern steganography uses the opportunity of hiding information into digital multimedia files and also at the network packet level.

Hiding information into a media requires following elements [2]
- The cover media(*C*) that will hold the hidden data
- The secret message (*M*), may be plain text, cipher text or any type of data
- The stego function (*Fe)* and its inverse (*Fe$^{-1}$*)
- An optional stego-key (*K*) or password may be used to hide and unhide the message.

The stego function operates over cover media and the message (to be hidden) along with a stego-key (optionally) to produce a stego media (*S*).
The schematic of steganographic operation is shown below.

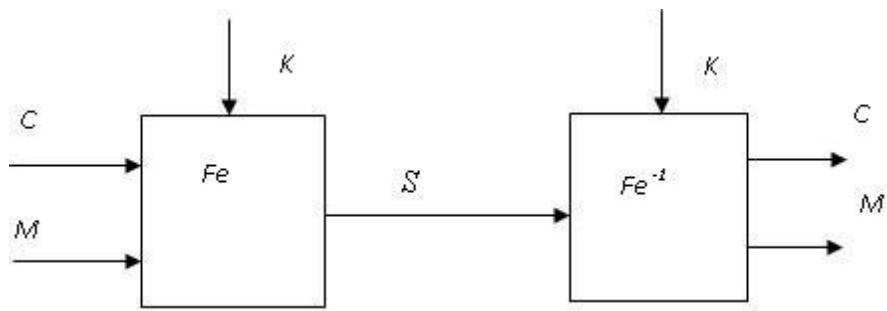

*Figure 1: The Steganographic operation*

Steganography and Cryptography are great partners in spite of functional difference. It is common practice to use cryptography with steganography.

## 2. Modern techniques of steganography

The common modern technique of steganography exploits the property of the media itself to convey a message.

The following media are the candidate for digitally embedding message [3]: -

- Plaintext
- Still imagery
- Audio and Video
- IP datagram.

### 2.1 Plaintext steganography

In this technique the message is hidden within a plain text file using different schemes like use of selected characters, extra white spaces of the cover text etc.

- **Use of selected characters of cover Text.**

    Sender sends a series of integer number (Key) to the recipient with a prior agreement that the secret message is hidden within the respective position of subsequent words of the cover text. For example the series is '1, 1, 2, 3, 4, 2, 4,' and the cover text is **"A team of five men joined today"**. So the hidden message is **"Atfvoa"**. A "0" in the number series will indicate a blank space in the recovered message. The word in the received cover text will be skipped if the number of characters in that word is less than the respective number in the series (Key) which shall also be skipped during the process of message unhide.

- **Use of extra white space characters of cover text.**

    A number of extra blank spaces are inserted between consecutive words of cover text. This numbers are mapped to a hidden message through an index of a

lookup table. For example extra three spaces between adjacent words indicate the number "3" which subsequently indicates a specific text of a look-up table which is available to the both communicating parties as a prior agreement.

**2.2 Still imagery steganography**

The most widely used technique today is hiding of secret messages into a digital image. This steganography technique exploits the weakness of the human visual system (HVS). HVS cannot detect the variation in luminance of color vectors at higher frequency side of the visual spectrum. A picture can be represented by a collection of color pixels. The individual pixels can be represented by their optical characteristics like 'brightness', 'chroma' etc. Each of these characteristics can be digitally expressed in terms of 1s and 0s.

For example: a 24-bit bitmap will have 8 bits, representing each of the three-color values (red, green, and blue) at each pixel. If we consider just the blue there will be $2^8$ different values of blue. The difference between 11111111 and 11111110 in the value for blue intensity is likely to be undetectable by the human eye. Hence, if the terminal recipient of the data is nothing but human visual system (HVS) then the Least Significant Bit (LSB) can be used for something else other than color information.

This technique can be directly applied on digital image in bitmap format as well as for the compressed image format like JPEG. In JPEG format, each pixel of the image is digitally coded using discrete cosine transformation (DCT). The LSB of encoded DCT components can be used as the carriers of the hidden message.

The details of above techniques are explained below:

•**Modification of LSB of a cover image in 'bitmap' format** [1]**.**

In this method binary equivalent of the message (to be hidden) is distributed among the LSBs of each pixel. For example we will try to hide the character 'A' into an 8-bit color image.
We are taking eight consecutive pixels from top left corner of the image. The equivalent binary bit pattern of those pixels may be like this: -

**00100111   11101001   11001000     00100111   11001000   11101001 11001000 00100111**

Then each bit of binary equivalence of letter 'A' i.e. **01100101** are copied serially (from the left hand side) to the LSB's of equivalent binary pattern of pixels, resulting the bit pattern will become like this: -

**0010011<span style="color:red">0</span>   1110100<span style="color:red">1</span>   1100100<span style="color:red">1</span>   0010011<span style="color:red">0</span>   1100100<span style="color:red">0</span>   1110100<span style="color:red">1</span> 1100100<span style="color:red">0</span> 0010011<span style="color:red">1</span>**

The only problem with this technique is that it is very vulnerable to attacks such as image compression and formatting.

- **Apply of LSB technique during discrete cosine transformation (DCT) [4] on cover image.**

The following steps are followed in this case: -

1. The Image is broken into data units each of them consists of 8 x 8 block of pixels.
2. Working from top-left to bottom-right of the cover image, DCT is applied to each pixel of each data unit.
3. After applying DCT, one DCT Coefficient is generated for each pixel in data unit.
4. Each DCT coefficient is then quantized against a reference quantization table.
5. The LSB of binary equivalent the quantized DCT coefficient can be replaced by a bit from secret message.
6. Encoding is then applied to each modified quantized DCT coefficient to produce compressed Stego Image.

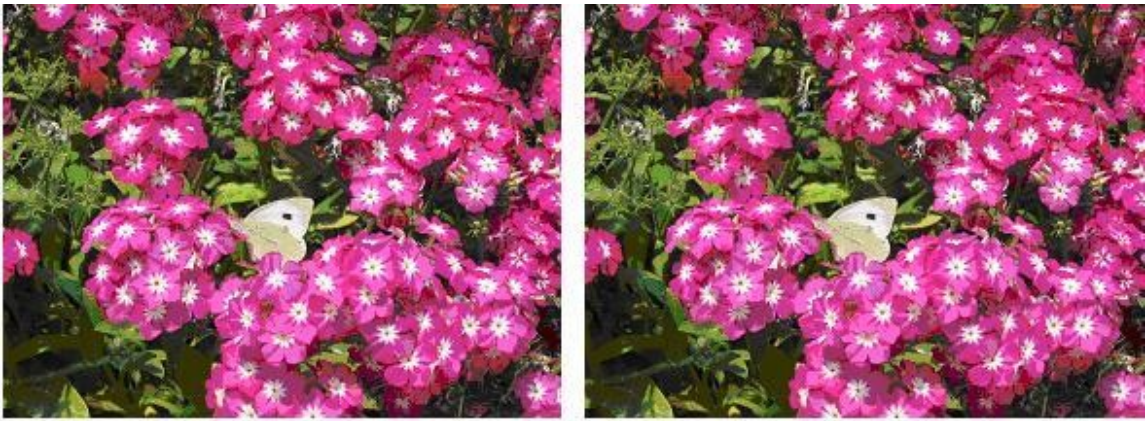

*Figure 2: Example of still imagery steganography. Left hand side image is the original cover image, where as right hand side does embedding a text file into the cover image make the stego image.*

## 2.3 Audio and Video Steganography [6, 7, 8]

In audio steganography, secret message is embedded into digitized audio signal which result slight altering of binary sequence of the corresponding audio file.
There are several methods are available for audio steganography. Some of them are as follows: -

- ➤ **LSB Coding:** Sampling technique followed by Quantization converts analog audio signal to digital binary sequence.

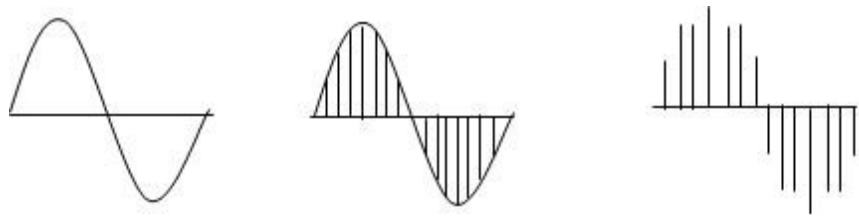

*Figure 3: Sampling of the Sine Wave followed by Quantization process.*

In this technique LSB of binary sequence of each sample of digitized audio file is replaced with binary equivalent of secret message.

For example if we want to hide the letter 'A' (binary equivalent **01100101**) to an digitized audio file where each sample is represented with 16 bits, then LSB of 8 consecutive samples (each of 16 bit size) is replaced with each bit of binary equivalent of the letter 'A'.

| Sampled Audio Stream (16 bit) | 'A' in binary | Audio stream with encoded message |
|---|---|---|
| 1001 1000  0011 1100 | 0 | 1001 1000  0011 110**0** |
| 1101 1011  0011 1000 | 1 | 1101 1011  0011 100**1** |
| 1011 1100  0011 1101 | 1 | 1011 1100  0011 110**1** |
| 1011 1111  0011 1100 | 0 | 1011 1111  0011 110**0** |
| 1011 1010  0111 1111 | 0 | 1011 1010  0111 111**0** |
| 1111 1000  0011 1100 | 1 | 1111 1000  0011 110**1** |
| 1101 1100  0111 1000 | 0 | 1101 1100  0111 100**0** |
| 1000 1000  0001 1111 | 1 | 1000 1000  0001 111**1** |

- ➢ **Phase Coding:**

Human Auditory System (HAS) can't recognize the phase change in audio signal as easy it can recognize noise in the signal. The phase coding method exploits this fact. This technique encodes the secret message bits as phase shifts in the phase spectrum of a digital signal, achieving an inaudible encoding in terms of signal-to- noise ratio.

- ➢ **Spread Spectrum** [17] **:**

There are two approaches are used in this technique: the direct sequence spread spectrum (DSSS) and frequency hopping spread spectrum (FHSS). Direct-sequence spread spectrum (DSSS) is a modulation technique used in telecommunication. As with other spread spectrum technologies, the transmitted signal takes up more bandwidth than the information signal that is being modulated.
Direct-sequence spread-spectrum transmissions multiply the data being transmitted by a "noise" signal. This noise signal is a pseudorandom sequence of 1 and −1 values, at a frequency much higher than that of the original signal, thereby spreading the energy of the original signal into a much wider band.

The resulting signal resembles white noise. However, this noise-like signal can be used to exactly reconstruct the original data at the receiving end, by multiplying it by the same pseudorandom sequence (because 1 × 1 = 1, and −1 × −1 = 1). This process, known as "de-spreading", mathematically constitutes a correlation of the transmitted Pseudo-random Noise (PN) sequence with the receiver's assumed sequence.

For de-spreading to work correctly, the transmit and receive sequences must be synchronized. This requires the receiver to synchronize its sequence with the transmitter's sequence via some sort of timing search process.

In contrast, frequency-hopping spread spectrum pseudo-randomly re-tunes the carrier, instead of adding pseudo-random noise to the data, which results in a uniform frequency distribution whose width is determined by the output range of the pseudo-random number generator.

> **Echo Hiding:**

In this method the secret message is embedded into cover audio signal as an echo. Three parameters of the echo of the cover signal namely amplitude, decay rate and offset from original signal are varied to represent encoded secret binary message. They are set below to the threshold of Human Auditory System (HAS) so that echo can't be easily resolved.

Video files are generally consists of images and sounds, so most of the relevant techniques for hiding data into images and audio are also applicable to video media. In the case of Video steganography sender sends the secret message to the recipient using a video sequence as cover media. Optional secret key 'K' can also be used during embedding the secret message to the cover media to produce 'stego-video'.

After that the stego-video is communicated over public channel to the receiver. At the receiving end, receiver uses the secret key along with the extracting algorithm to extract the secret message from the stego-object.

The original cover video consists of frames represented by $C_k(m,n)$ where $1 \leq k \leq N$. 'N' is the total number of frame and m,n are the row and column indices of the pixels, respectively. The binary secret message denoted by $M_k(m, n)$ is embedded into the cover video media by modulating it into a signal. $M_k(m, n)$ is defined over the same domain as the host $C_k(m, n)$. The stego-video signal is represented by the equation

$S_k(m, n) = C_k(m, n) + \alpha_k(m, n) M_k(m, n)$ ,  k = 1, 2, 3 . . . N

where $\alpha_k(m, n)$ is a scaling factor. For simplicity $\alpha_k(m, n)$ can be considered to be constant over all the pixels and frames.
So the equation becomes:

$S_k(m, n) = C_k(m, n) + \alpha(m, n) M_k(m, n)$ ,  k = 1, 2, 3 . . . N

## 2.4 IP datagram steganography [5, 9]

This is another approach of steganography, which employs hiding data in the network datagram level in a TCP/IP based network like Internet. Network Covert Channel is the synonym of network steganography. Overall goal of this approach to make the stego datagram is undetectable by Network watchers like sniffer, Intrusion Detection System (IDS) etc.

In this approach information to be hide is placed in the IP header of a TCP/IP datagram. Some of the fields of IP header and TCP header in an IPv4 network are chosen for data hiding.

First we will demonstrate how 'Flags' and 'Identification' field of Ipv4 header can be exploited by this methodology.

*Figure 4: IPv4 header*

- **Covert Channel Communication using 'Flags' field:**

The size of Flag field is 3 bit. There are 3 flags denoted by each bit. First bit is reserved. Second and third one denoted by DF (Don't fragment) and MF (More Fragment) respectively. An un-fragmented datagram has all zero fragmentation information (i.e. MF = 0 and 13-bit Fragment Offset = 0) which gives rise to a redundancy condition, i.e. DF (Do not Fragment) can carry either "0" or "1" subject to the knowledge of the maximum size of the datagram.

Now if sender and recipient both have a prior knowledge of Maximum Transfer Unit (MTU) of their network then they can covertly communicate with each other using DF flag bit of IP header. Datagram length should be less than path MTU otherwise packet will be fragmented and this method will not work.

The following table shows the how the sender communicates 1 and 0 to the recipient by using DF flag bit.

| Datagram | 3-bit Flag field | 13-bit fragment offset | Remarks |
|---|---|---|---|
| 1 | 0**1**0 | 00…00 | Datagram 1 covertly communicating **'1'** |
| 2 | 0**0**0 | 00…00 | Datagram 2 covertly communicating **'0'** |

This is an example of covert communication since there is no way to the network monitoring devices like IDS or sniffer to detect the communication because cover datagram is a normal datagram. As the payload is untouched, there is no way an IDS or any other content filtering device could recognize this activity. In major constraint of this approach is both parties should have prior knowledge of path MTU and datagram from sender should not be fragmented further in the way.

- **Covert Channel Communication using 'Identification' field:**

    The '16-bit identification field' in Ipv4 header[10] is used to identify the fragmented packed of an IP datagram. If there is no fragmentation of datagram, then this Identification field can be used to embed sender specified information.

- **Covert Channel Communication using ISN (Initial Sequence Number) field:**

    ISN (Initial Sequence Number) in TCP header[14] is another candidate for cover media for network steganography. Initial sequence number is a 32-bit digit generated during three-way TCP/IP handshaking between client and server, which is as follows: -
    (a) Client sends a TCP/IP packet with SYN flag is ON. This is segment 1 where client specifies the port number of the server that it wants to connect to, and the client's ISN.
    (b) Sever responds with a TCP/IP packet with SYN flag is ON and also containing the server's ISN. Server also acknowledges the client's SYN by ACK flag is ON in this packet with acknowledgement number which is client's ISN+1. This is segment 2.
    (c) The client must acknowledge this server's SYN packet by sending a packet with ACK flag is ON with acknowledgement number, which is server's ISN+1. This is segment 3.

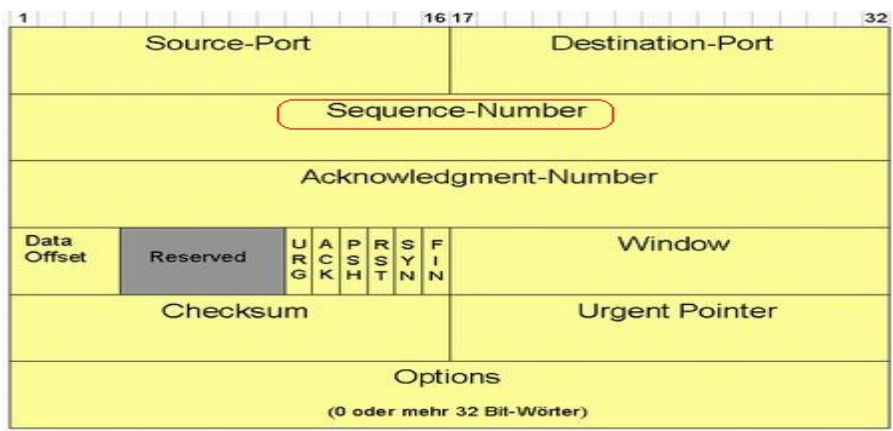

*Figure 5: TCPv4 header*

The large 32-bit address space of the Sequence Number field can be used for covert channel. The sender will craft a TCP/IP packet where the secret binary message can be embedded over the Sequence Number field and the passively listening receiving party will then extract the data.

Source Port and Checksum in UDP header[15] Code field in ICMP header[13] are also good candidate for Cover item for Network Steganography.

## 3. Steganalysis [16]

Steganalysis is the process of identifying steganography by inspecting various parameter of a stego media. The primary step of this process is to identify a suspected stego media. After that steganalysis process determines whether that media contains hidden message or not and then try to recover the message from it.

In the cryptanalysis it is clear that the intercepted message is encrypted and it certainly contains the hidden message because the message is scrambled. But in the case of steganalysis this may not be true. The suspected media may or may not be with hidden message. The steganalysis process starts with a set of suspected information streams. Then the set is reduced with the help of advance statistical methods.

- **Steganalysis Techniques**

The properties of electronic media are being changed after hiding any object into that. This can result in the form of degradation in terms of quality or unusual characteristics of the media: Steganalysis techniques based on unusual pattern in the media or Visual Detection of the same.

For example in the case of Network Steganography unusual pattern is introduced in the TCP/IP packet header. If the packet analysis technique of Intrusion Detection System of a network is based on white list pattern (usual pattern), then this method of network steganography can be defeated.

In the case of Visual detection steganalysis technique a set of stego images are compared with original cover images and note the visible difference. Signature of the hidden message can be derived by comparing numerous images. Cropping or padding of image also is a visual clue of hidden message because some stego tool is cropping or padding blank spaces to fit the stego image into fixed size. Difference in file size between cover image and stego images, increase or decrease of unique colors in stego images can also be used in the Visual Detection steganalysis technique.

- **Steganography Attacks**

Steganographic attacks consist of detecting, extracting and destroying hidden object of the stego media. Steganography attack is followed by steganalysis. There are several types of attacks based on the information available for analysis. Some of them are as follows: -

- **Known carrier attack:** The original cover media and stego media both are available for analysis.
- **Steganography only attack:** In this type of attacks, only stego media is available for analysis.
- **Known message attack:** The hidden message is known in this case.
- **Known steganography attack:** The cover media, stego media as well as the steganography tool or algorithm, are known.

## 4. Conclusion

In this paper, different techniques are discussed for embedding data in text, image, audio/video signals and IP datagram as cover media. All the proposed methods have some limitations. The stego multimedia produced by mentioned methods for multimedia steganography are more or less vulnerable to attack like media formatting, compression etc. In this respect, IP datagram steganography technique is not susceptible to that type of attacks. Steganalyis is the technique to detect steganography or defeat steganography. The research to device strong steganographic and steganalysis technique is a continuous process.